# Fabrication and characterization of iodine photonic microcell for sub-Doppler spectroscopy and laser stabilization

CLÉMENT GOÏCOECHÉA,[1,2] THOMAS BILLOTTE,[2] MATTHIEU CHAFER,[1,2] MARTIN MAUREL,[1,2] JENNY JOUIN,[3] PHILIPPE THOMAS,[3] DEVANG NAIK,[2] FRÉDÉRIC GÉRÔME,[1,2] BENOÎT DEBORD,[1,2] AND FETAH BENABID[1,2,*]

[1]*GLOphotonics SAS, 123 avenue Albert Thomas, 87060 Limoges, France*
[2]*GPPMM group, Xlim Research Institute, CNRS UMR 7252, University of Limoges, Limoges, France*
[3]*IRCER UMR CNRS 7315, Centre Européen de la Céramique, 12 rue Atlantis, 87068 Limoges, France*
*[\*]f.benabid@xlim.fr*

**Abstract:** We report on the development of all-fiber stand-alone Iodine-filled Photonic Microcells demonstrating record absorption contrast at room temperature. The microcell's fiber is made of inhibited coupling guiding hollow-core photonic crystal fibers. The fiber-core loading with Iodine was undertaken at $10^{-1}$-$10^{-2}$mbar vapor pressure using a novel gas-manifold based on metallic vacuum parts with ceramic coated inner surfaces for corrosion resistance. The fiber is then sealed on the tips and mounted on FC/APC connectors for better integration with standard fiber components. The stand-alone microcells display Doppler lines with contrasts up to 73% in the 633nm wavelength range, and an insertion loss between 3 to 4dB. Sub-Doppler spectroscopy based on saturable absorption has been carried out to resolve the hyperfine structure of the P(33)6-3 lines at room temperature with a full-width at half maximum of 24MHz on the b4 component with the help of lock-in amplification. Also, we demonstrate distinguishable hyperfine components on the R(39)6-3 line at room temperature without any recourse to signal-to-noise ratio amplification techniques.

## 1. Introduction

In the last 20 years, efforts have been made towards the miniaturization of frequency standards with the emergence of new technological devices, such as clocks based on microelectromechanical systems (MEMS) [1], planar devices mounted on silicon chip, using hollow-core anti-resonant reflecting optical waveguides (ARROW) [2], or compact engineered circular multi-pass cells [3]. Amongst these devices, the hollow-core photonic crystal fiber (HCPCF) technology appeared to be an excellent and promising alternative to bulky cells for portable, small footprint applications by filling and sealing atomic or molecular vapor inside its core [4], culminating in the form of the photonic microcell (PMC) an atom-photonics component that can be integrated with small insertion loss while using existing fiber connectors into any optical set-up [5]. By confining the atoms/molecules alongside light over modal areas of as small as a few μm$^2$, whilst keeping them in interaction over length scales a million times longer than the Rayleigh range, the resulting enhanced atom-laser interaction efficiency leads to strong absorption despite low gas density and low light-level, resulting in an increased signal to noise ratio compared to other technologies. The unique optical properties of HCPCF offer a large range of core sizes and lengths coupled with low loss, alongside reduced power consumption and micro-structured cladding providing versatile modes composition allowing transverse light structuring [6].

HCPCFs light guiding performances are continuously improving on a broad spectral range with loss figures competing with standard solid-core fibers on telecom spectral ranges with loss down to 0.174dB/km in the C-band [7]. Low-loss figures are also accessible in the visible range down to 0.9dB/km at 558nm [8].

Since the advent of the PMC as a photonic component, a plethora of work has been carried out on the types of enclosed gas and the sealing techniques. The evolution of the different fabrication techniques of PMC has been mainly dictated by the type of fiber as gas-container and by the nature of the gas. In fact, compatibility of the solid-core single mode fiber (SMF) splicing technique with photonic bandgap (PBG) fibers [9,10] with its associated core diameter range of 5-20µm is no longer viable with inhibited-coupling guiding HCPCF (IC-HCPCF) because of their larger core sizes (typically between 20 and 100 µm), and thus implies a strong mode mismatch induced loss. Therefore, mode-field adapters have been introduced to render IC guiding fiber technology compatible with SMF, either by tapering HCPCF as reported by Wheeler *et al*. [11] or by implementing graded-index fibers [12,13]. Another configuration has been also reported based on glass cells glued on the tips of PBG HCPCF to manage gas filling and proceed to gas-enclosing by collapsing a section of the glass cell [14]. All the different techniques mentioned above suffer from drawbacks linked to the use of glue or exposition to ambient air leading to gas contamination and degradation of PMC spectroscopic performances on the long term. Recently, Billotte *et al.* introduced a novel PMC assembly process that no longer requires helium buffer gas or gluing stage [5]. Based on a glass sleeve collapse, a 1.5dB insertion loss microcell has been achieved with a 7m long IC-HCPCF filled with acetylene at 80µbar pressure. Doppler-free spectroscopy showed constant linewidth and contrast over more than 3 months, thus highlighting the quality/purity of the sealed-gas medium and the impact of this contaminant-free technique for the creation of next-generation PMCs.

The above work on PMC assembly was chiefly motivated for frequency standards, which thus far was limited to molecular gases, such as Acetylene ($C_2H_2$, 1550nm region) [4,5,11] or Carbon Dioxide ($CO_2$) [15–17]. The actual state-of-the-art for a sealed PMC is set by Triches *et al.* and Billotte *et al.*, each exhibiting an instability around $2.10^{-11}$ @1000s [14,18]. Both are working with acetylene around the telecom wavelength band.

Extending PMC technology to atomic or molecular vapors, such as alkali vapors or molecular halogen gas (such as iodine ($I_2$)), have been very challenging because of their physio-chemical reactivity and the required vacuum environment. This in turn limited the impact of PMC based optical frequency references to a restricted number of spectral ranges. For example, in the green and red spectral range, $I_2$ is known for displaying a very dense Doppler-lines spectrum useful for the development of visible broadband frequency references [19,20]. This is illustrated by the fact that several $I_2$ ro-vibrational absorption lines are among Bureau International des Poids et Mesures (BIPM) recommended frequency references for the realization of the meter [21]. Also, $I_2$ based frequency stabilization [22–26] proved to be valuable for several technological applications [27–30]. Consequently, the advent of iodine filled PMC ($I_2$-PMC) would be extremely useful for applications like guiding star lasers [31], high resolution LIDAR [32], or laser frequency stabilization [26], which requires these performances to be delivered in a compact and mobile physical package. However, encapsulating iodine vapor into glass cells carries specific challenges because of $I_2$ physical and chemical properties. In particular, its high reactivity and corrosion with metals [33,34] requires the use of complex glass-manifold filling system. Consequently, the development of $I_2$-PMC represents a technical challenge both in vapor handling, loading and in-HCPCF sealing. Indeed, the use of common metallic vacuum parts is inadequate for the $I_2$-PMC assembly process, and the commonly used glass manifold alternative for $I_2$ cell manufacturing is too complex and expensive for scalable $I_2$-PMC fabrication. Within this context, we note the previous work on $I_2$-filled HCPCF [24,35,36]. Lurie *et al.* have shown that the hollow-core fiber for $I_2$-microcell development and saturated absorption spectroscopy applications demonstrated a strong efficiency thanks to strong overlap of pump and probe beams [35], and laser stabilization with fractional frequency stability of $2.3\times10^{-12}$ at 1s for a HCPCF mounted on a glass vacuum manifold was achieved [24]. Impact of residual gases in the vacuum system alongside $I_2$ pressure have been raised by the authors as obstacles to reach transit limited sub-Doppler linewidths. This seminal work within the $I_2$-filled HCPCF framework has been followed up in 2015 by the first demonstration of a hermetically sealed $I_2$ Kagome HCPCF [36]. However, despite a demonstration of laser stabilization with fractional frequency stabilities of $3.10^{-11}$ at 100s, the sealed HCPCF has been marred by a strong 21.5dB insertion loss and cannot be coined PMC because the gas sealing was achieved by fusion-collapsing a section of the fiber, thus eliminating its optical guidance features at the collapsed section.

We report in this work on an $I_2$-PMC development based on a scalable, corrosion resistant and contamination free fabrication process [5]. For example, a 2.5 and 4 meter long patch-cord like iodine PMC based on tubular IC-HCPCF [37] and an overall transmission efficiency as high as 40% have been fabricated. These PMCs exhibit, at room temperature, high performances in term of absorption contrast reaching respectively 60% and 53% on the P(33)6-3 transition - at 632.991nm [38]. Furthermore, we demonstrate room temperature resolution of Iodine hyperfine spectrum observation, over 10GHz spectral range around 633nm, of several sub-Doppler spectral transparencies from different broadened lines, thanks to saturable absorption. The exceptional lifetime of these microcells is demonstrated through the unaffected P(33)6-3 absorption contrast and Doppler linewidth of 4 year-old $I_2$-PMC.

## 2. Experimental set-up for $I_2$-PMC fabrication

An IC-HCPCF based on tubular lattice cladding has been designed and fabricated for optimal guidance on several hundred thousand hyperfine transitions of iodine in the green-to-red spectral range (Fig. 1(a)) with loss below 30dB/km level between 530nm and 668nm. The fiber (Fig. 1(a)) with an outer diameter of 200µm exhibits a core diameter of 30µm surrounded by 8 isolated tubes cladding. This fiber presents an excellent modal behavior with quasi single-mode guidance as illustrated by the near-field intensity profile at 633nm (see Fig. 1(a)) measured at the output of a 4m long fiber.

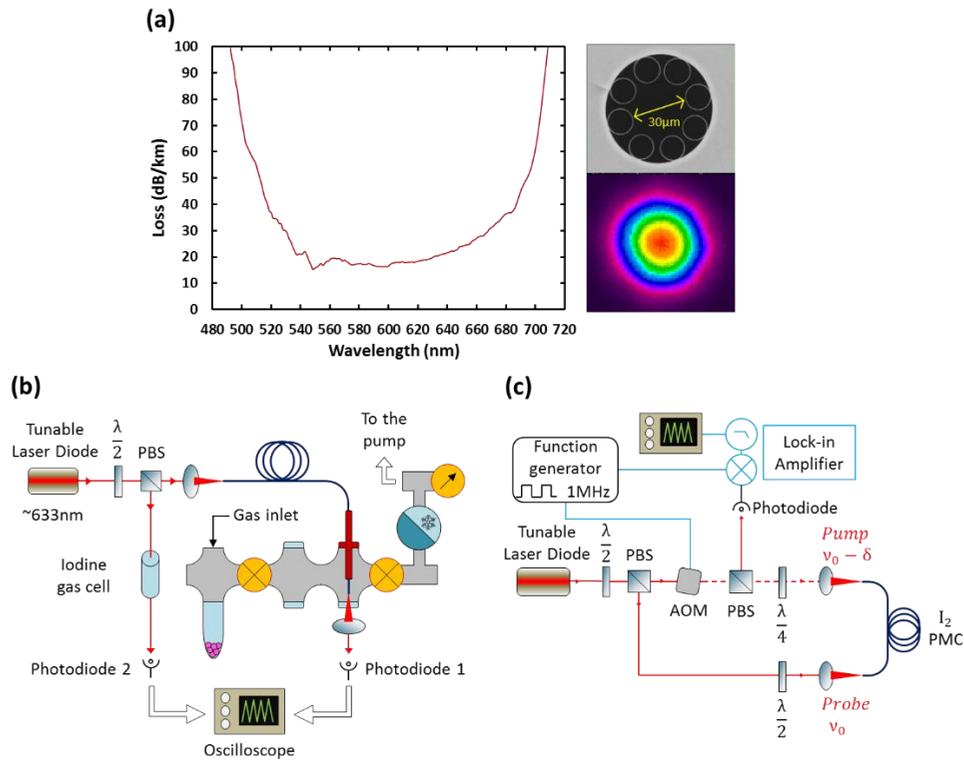

Fig. 1. (a) Measured loss spectrum of the experimental IC tubular fiber related to the developed PMCs (see Fig. 2). On the right : micrograph picture of the fiber cross section and near field intensity distribution at 633nm. (b) Overview of the experimental set-up for $I_2$-filling and fiber-sealing. The vacuum system is represented in gray with valves and the gauge represented by yellow circles. The fiber is represented in dark blue. PBS: polarizing beam splitter. Lock-in system is represented in blue color. AOM : acousto-optic modulator. (c) Schematic of set-up for PMC characterization.

Two PMCs, PMC#1 and PMC#2, have been fabricated 3 years apart in 2017 and 2020 respectively from similar fibers using an in-house gas-vacuum manifold, represented schematically in Fig. 1(b) and purposely designed for $I_2$ loading into HCPCF and for $I_2$-PMC assembly.

The manifold is composed of three main compartments separated by vacuum valves. The central part holds the HCPCF and acts as the fiber loading section. On the right side of the fiber loading section, a turbo-molecular vacuum-pump is connected via a cryogenic trap to prevent any contamination to the vacuum-pump during $I_2$ releasing. The left side corresponds to the $I_2$ dispenser. Here, iodine chips are placed in glass test-tube, which is hermetically connected to the manifold via a metallic fitting. This section is under pressure and/or temperature regulation for $I_2$ sublimation and release into the fiber loading section. The fiber loading goes through the following sequence. First, the fiber loading section is evacuated to a vacuum pressure of less than $10^{-6}$ mbar. Similarly, the iodine dispenser section is evacuated while ensuring the $I_2$ remains solid by regulating the iodine chip temperature. The above ensures the leakproofness of the manifold and high vacuum quality. Once this process is achieved, the $I_2$ is sublimated by increasing the temperature until a pressure of around $10^{-1}$ mbar is reached. Second, the $I_2$ vapor is released in the fiber loading section by opening the valve between the two sections and closing the valve to the vacuum pump section. During this loading process, we continuously monitor the fiber transmission spectrum derived from a tunable external cavity diode laser (TOPTICA DL-PRO, 631-635nm range) tuned to a frequency range corresponding to one of the iodine rovibrational lines. A second beam from the same laser is sent to an Iodine macroscopic gas cell and serves as a reference. The spectroscopic signature of the Iodine absorption lines (spanning over a set of 3 lines around the P(33)6-3 transition) can be observed after 10 minutes of loading. The loading is then kept on until the desired contrast is reached.

It is noteworthy, that the metallic parts of the whole manifold have been post-processed against corrosion and chemical reaction with iodine by applying a ceramic coating on the inner metallic surfaces. This allows an outstanding ease-of-use with several HCPCFs being loaded and assembled over several years.

Before mounting and splicing the HCPCF (described in Fig. 1(a)), it was flushed with Helium or Argon gas and heated for several hours in the oven at ~100°C to reduce any residual gas inside the fiber. The fiber is then end-capped on one extremity by collapsing a borosilicate capillary with an inner diameter fitting the outer diameter of the fiber, following the process mentioned in [5]. The sealed and polished extremity is then mounted on a FC/APC optical connector with a measured coupling loss in the range of 1 to 1.5dB at 633nm, 20dB lower than the splicing loss obtained by collapsing the fiber on itself in [36].

The second extremity of the HCPCF is connected by borosilicate sleeve fusion splicing to a 30cm piece of the same HCPCF. The second tip of 30cm long HCPCF is hermetically attached to the loading compartment of the manifold via a home-made fiber-feedthrough (Fig. 1(a) of [5]). The end-capped fiber is then evacuated by pumping the valve-controlled middle chamber of the vacuum system down to the range of $10^{-6}$ mbar. Once the desired contrast is reached (here around 60%), we hermetically seal the fiber by end-capping with a splicing machine based on sleeve collapse around the tip of the HCPCF, as described in [5]. The tips of the resulted PMC are then polished and mounted on FC/APC fiber connectors. Figure 2(a) shows the photography of a typical FC/APC connectorized patch-cord like PMC in its final form. Figure 2(b) shows the reconstructed near-field intensity profile of the transmitted light from the developed $I_2$-PMC. The measured transmission was in the range of 40-50%, corresponding to an insertion loss of less than 4 dB, which is 17.5dB lower than the one measured in [36].

Figure 2(d) represents the normalized transmission spectra at the output of PMC#1 (red curve), PMC#2 (orange curve) and the commercial macroscopic cell (black curve) measured consecutively with the same laser configurations and room temperature conditions. The bottom axis and the top axis of the graph give the frequency and the relative frequency from that of the R(39)6-3 transition, respectively. The two PMCs exhibit contrast between 53% to 60% for the P(33)6-3 line and 61% to 73% for the R(39)6-3. This is respectfully 5.9 to 6.7 and 5.1 to 6.1 times larger than the ones obtained with the 10cm long commercial $I_2$ macroscopic gas cell (resp. 9% & 12%) corresponding to $1.3.10^{-1} - 1.6.10^1$ mbar of $I_2$ vapor pressure specifications given by manufacturer. These contrasts are 2.2 to 2.9 times larger than the one obtained at room temperature in previously reported work [36], which is to our knowledge the only reported work on low-loss $I_2$-loaded sealed HCPCF.

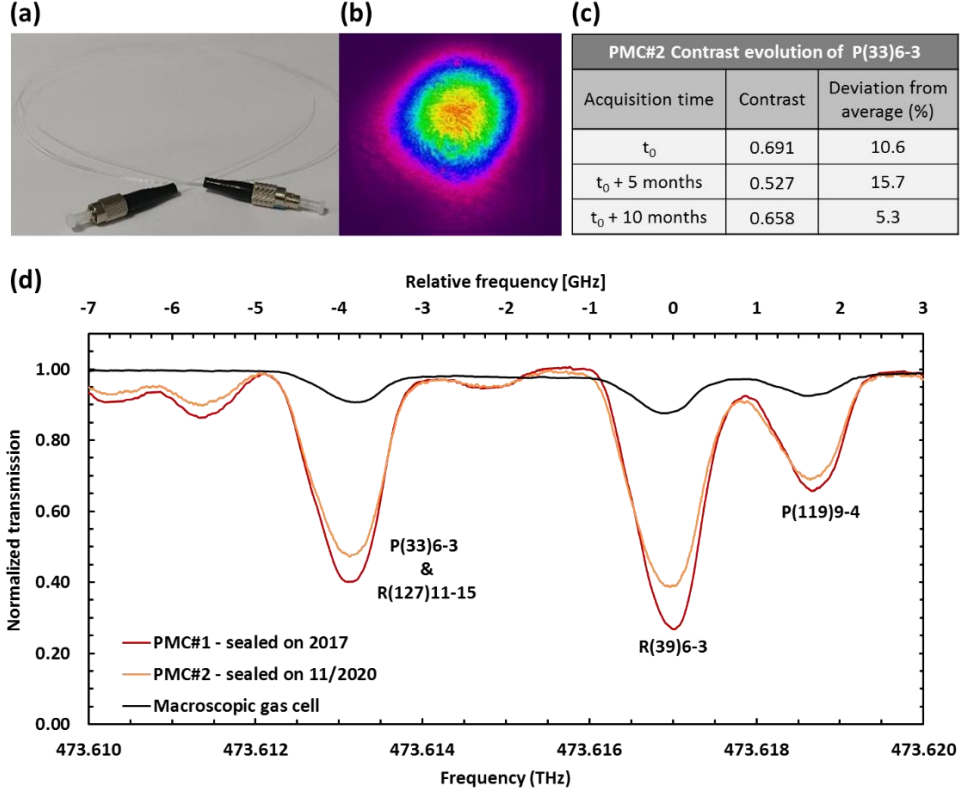

Fig. 2. (a) Photography of I$_2$ PMC#1 mounted on FC/APC connectors. (b) Measured near field intensity profile at 633nm at the output of PMC#1. (c) Contrast evolution of the PMC#2 over 2 years. (d) Normalized transmission spectra through the fabricated PMCs (PMC#1 in red color, PMC#2 in orange) and through a macroscopic commercial gas cell (black).

Finally, comparison of the shown transmission spectra with those recorded at the time of PMC sealing and more recently on October 2022 (see Fig.2(c)) shows comparable contrasts. In fact, evolution of the contrast of P(33)6-3 line through PMC#2 has been studied along 2 years since its encapsulation at $t_0$. The different measurements are summarized on table from Fig. 2(c). The measured contrast of the P(33)6-3 line of I$_2$ was found to remain constant within a range of ±8,5% around the extrema average value of 61%. Observed fluctuations have been attributed to the different temperature conditions and laser diode ampereage setpoint, and corroborated by an additional study using another PMC (based on the same fiber and fabrication process). The result has shown that by considering the extreme experimental values of room temperature (*i.e.* from 19 to 22.5°C) and laser diode ampereage, these two major contributions of contrast change can lead to a variation of ±9%.

The stability of the absorption contrast highlights the leak proofness of the PMCs, and the reliability and repeatability of the developed process.

## 3. Sub-Doppler spectroscopy with I$_2$ PMC

The fabricated PMCs have shown their potential for sub-Doppler spectroscopy through the resolution of the hyperfine structure of the P(33)6-3 line. To do so, Saturated Absorption

Spectroscopy (SAS) measurements have been done following the set-up shown in Fig. 1(c). The laser beam is separated into counter-propagating pump and probe beams with 4mW and 40μW output power, respectively. The pump beam is obtained from the first order diffracted beam off an acousto-optic modulator (AOM) operating at 64MHz frequency. This was motivated so to avoid interference between the probe and back-reflected pump during the propagation in the fiber. The spectroscopic transparency signal is obtained by redirecting the PMC-transmitted probe beam on to a photodiode with the help of a polarizing cube. Half and quarter waveplates are used to improve both the optical PMC transmission and the intensity of the redirected probe beam on the photodetector.

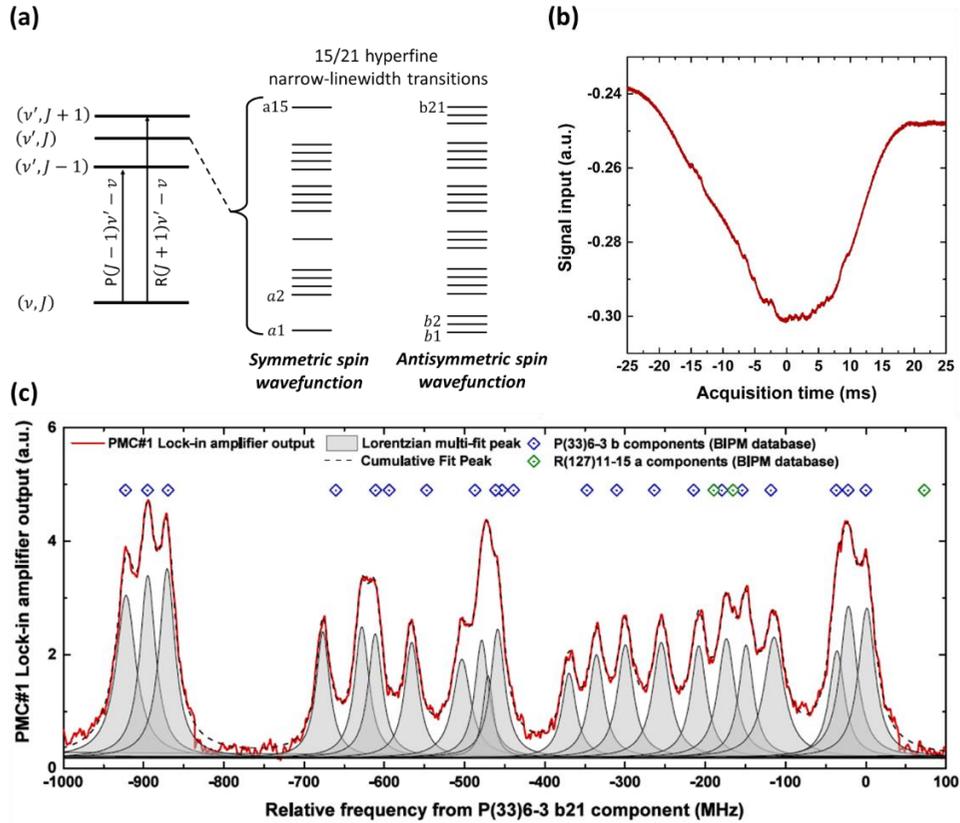

Fig. 3. (a) Example of Iodine ro-vibrational hyperfine energy levels. *This structure is usually not observable through a simple macroscopic cell at room temperature*. (b) Measured R(39)6-3 Doppler line at room temperature through PMC#1. (c) Hyperfine components structure of the P(33)6-3 Doppler line obtained with a lock-in amplifier detection scheme. Measured peaks (red) are compared with tabulated values components for P(33)6-3 & R(127)11-15 (respectively in blue and green). Data have been fitted with lorentzian multi-peak fit function.

Figure 3(b) shows the probe signal when the laser frequency is tuned in the vicinity of R(39)6-3 line and recorded directly by the photodetector at room temperature. In addition of the Doppler-broadened absorption line, the trace shows the 21 hyperfine b-lines [39]. To our knowledge, this is the first time that such Iodine transparencies are observed on a cell without any means of signal-to-noise ratio (SNR) post-acquisition amplification such as lock-in detection. In order to improve the hyperfine structure resolution we used a lock-in amplification detection scheme with a squared-modulated pump beam of 1MHz (amplitude modulation). The spectrum obtained as output of the lock-in amplifier is shown in Fig. 3(c). The 21 hyperfine b-components of the P(33)6-3 line ("b" energy level scheme shown in Fig.3(a)) have been identified and are in good agreement with the optical frequencies tabulated by the BIPM in blue [38]. One can notice some shift and/or additional peak, such as between b11 and b18, that could be explained by the overlap between P(33)6-3 and R(127)11-5 absorption lines of $I_2$. Hence, additional weaker peaks could come from the SAS of R(127)11-5 line.

Figure 4 shows the b4 hyperfine component of the P(33)6-3 line, previously displayed in Fig. 3(c), on which a Lorentzian fitting shows a FWHM of 21MHz for 4mW pump beam. Contribution of broadening sources such as wall collisions and the natural linewidth can be directly calculated [4] leading to 2.95MHz and 3.23MHz respectively (lifetime about 310ns [40,41]). The laser linewidth provided by the manufacturer is about 0.20MHz. Considering the pressure of $I_2$ inside the PMC of around $10^{-2}$mbar, we can estimate a few 40kHz [42] intermolecular collision broadening. Therefore, the minimum linewidth obtainable using this setup is 6.42MHz. The larger measured linewidth of 21MHz is explained by power broadening coming from the pump beam intensity of 10MW/m², 357 times bigger than saturation intensity of 28kW/m² [43].

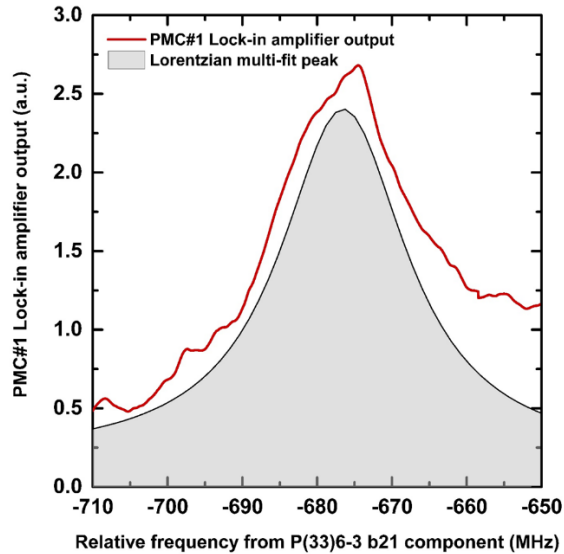

Fig. 4. Zoom-in on b4 component of the hyperfine structure displayed in Fig. 3(c). Lock-in signal is displayed in red line. A 21MHz FWHM Lorentzian curve has been fitted in black line.

## 4. Conclusion

As a summary, we reported on the first fabrication of meter-long low-optical loss pure $I_2$ PMCs based on a new process for creating all-fibered stand-alone PMCs. Absorption contrasts up to 73% have been measured at room temperature with PMC insertion loss of 4dB, 5.4 times lower than the state-of-the-art. The good sealing quality is demonstrated by a PMC being still functional after 4 years and the stable absorption measured throughout both PMCs, as well as the performance of Doppler-free signal measurements for the first PMC on the P(33)6-3 line of $I_2$ at room temperature. The b4 component of this line shows a FWHM of 24MHz at 4mW output pump beam power. By calculating the different broadening sources, we identify the dominant one as power broadening. An optimization of $I_2$ pressure, PMC length and pump power should allow us to reduce this FWHM while keeping the same SA contrast. These results are very promising for many compact sensing applications and laser stabilization.

**Funding:** Région Nouvelle-Aquitaine.

**Disclosures:** The authors declare no conflicts of interest.